\documentclass[twocolumn,longbibliography,superscriptaddress]{revtex4-2}

\usepackage[english]{babel}
\usepackage{physics}
\usepackage[dvipdfmx]{graphicx}
\usepackage{amsmath,amssymb}
\usepackage{xparse}
\usepackage{color}

\def\vk{\vb*{k}}

\def\ho{\hbar\omega}
\def\hO{\hbar\Omega}
\def\ta{t_{\mathrm{a}}}
\def\tr{t_{\mathrm{r}}}

\begin{document}

\title{Theory of Photocurrent and High-Harmonic Generation with Chiral Fermions}

\author{Yuya Ominato}
\affiliation{Waseda Institute for Advanced Study, Waseda University, Shinjuku-ku, Tokyo 169-0051, Japan}
\author{Masahito Mochizuki}
\affiliation{Department of Applied Physics, Waseda University, Okubo, Shinjuku-ku, Tokyo 169-8555, Japan}

\date{\today}

\begin{abstract}
We theoretically discover possible dc-current induction and high-harmonic generation from photodriven chiral fermions in B20-type semimetals irradiated with circularly polarized light as nonlinear optical responses with several unconventional properties. First, we find multiple sign changes of the induced bulk dc photocurrent as a function of light parameters, which is ascribed to the nature of asymmetric photon-dressed bands in chiral systems. Moreover, we observe a parity-dependent directivity of high-harmonic generation where the odd- and even-order harmonics have intensities only in directions perpendicular and parallel to the polarization plane, respectively, which can be understood from dynamical symmetry of the present photodriven chiral systems.
\end{abstract}

\maketitle 

\section{Introduction}
\label{sec:introduction}

The nonlinear optical response induced by high-intensity lasers is a central topic in modern condensed matter physics \cite{Bloembergen1996,Boyd2008,Nagaosa2022,Morimoto2023}.
In particular, understanding the mechanisms of the bulk photovoltaic effect and high-harmonic generation (HHG) is of great interest.  
The bulk photovoltaic effect generates dc photocurrent under irradiation and is allowed in noncentrosymmetric materials \cite{Von1981,Sturman1992,Sipe2000,Young2012a,Young2012b,Morimoto2016,Dai2023-fh}.
Recently, dc photocurrent generation in noncentrosymmetric Weyl semimetals has attracted significant attention \cite{Ma2017,Sun2017-br,Osterhoudt2019,Sirica2019,Wang2019,Ma2019,Gao2020}.
Additionally, since the first observation of HHG in semiconductors \cite{Ghimire2011}, HHG has been extensively studied in various solids \cite{Ghimire2019,Yue2022,Li2023,Bhattacharya2023,Hirori2024}.
HHG has been observed in topological materials, such as Weyl semimetals \cite{Patankar2018,Lv2021}, Dirac semimetals \cite{Cheng2020,Kovalev2020}, and topological insulators \cite{bai2021high,Schmid2021,Baykusheva2021}.
These studies contribute to the dynamic control of chiral fermions in solids and the exploration of quantum phenomena in strongly driven systems \cite{Oka2009,Oka2019,Rudner2020}.

B20-type chiral semimetals have long been studied in various contexts, for example, as thermoelectric materials \cite{Asanabe1964,Mcneill1964,Rowe1995}, and their significance as a platform for hosting unconventional chiral fermions \cite{Bradlyn2016,Rarita1941,Manes2012,Wieder2016,Weng2016a,Weng2016b,Zhu2016,Liang2016,Ezawa2016} has recently attracted growing interest.
Several theoretical studies have predicted the emergence of chiral multifold fermions in B20-type chiral semimetals, such as CoSi \cite{Tang2017,Pshenay-Severin2018} and RhSi \cite{Chang2017}, and recent experiments have demonstrated characteristic features of the chiral multifold fermions through optical responses \cite{Xu2020,Maulana2020} and angle-resolved photoemission spectroscopy \cite{Lv2017,Takane2019,Sanchez2019,Schroter2019,Rao2019,Li2019}.
Moreover, the circular photogalvanic effect, a hallmark of nonlinear optical responses with chiral fermions, has been observed \cite{Rees2020,Ni2020,Ni2021}.
It is a natural extension of interest to explore characteristic nonlinear optical responses of B20-type chiral semimetals induced by intense laser irradiation, where nonequilibrium effects become more pronounced.
Such investigations remain a promising avenue for further exploration \cite{Ezawa2017,Ahn2023-vo,Fan2024-no}.

\begin{figure}[b]
\begin{center}
\includegraphics[width=1\hsize]{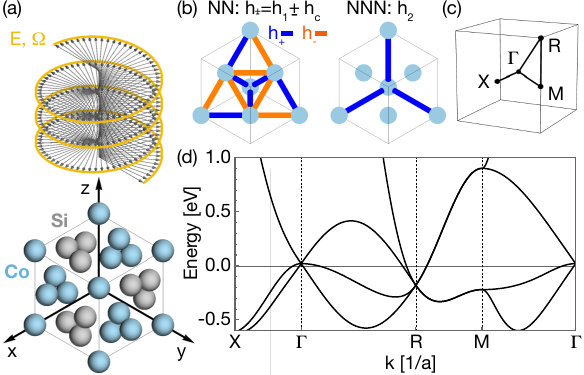}
\end{center}
\caption{
(a) Circularly polarized light applied along the $z$-axis to the B20-type chiral semimetal CoSi.
(b) Schematics of the nearest-neighbor (NN) and next-nearest-neighbor (NNN) transfer integrals of the tight-binding model.
(c) Brillouin zone and high-symmetry points.
(d) Energy band structure of the tight-binding model without light irradiation.}
\label{fig_system}
\end{figure}

In this paper, we propose that bulk harmonic photocurrent of arbitrary order, consisting of both dc and ac components, is generated in B20-type chiral semimetals by circularly polarized light irradiation.
We consider CoSi irradiated by circularly polarized light with electric field $E$ and frequency $\Omega$, as shown in Fig.~\ref{fig_system}(a).
Light propagates in the (001) direction, which is suitable for the symmetry analysis.
The dc-photocurrent generation corresponding to the circular photogalvanic effect manifests as a nonlinear optical response. The dc-photocurrent exhibits multiple sign changes as a function of light parameters, $E$ and $\Omega$.
Additionally, the ac harmonic photocurrent gives rise to HHG.
The HHG intensity exhibits parity-dependent directivity, which means odd- and even-order harmonics are directed perpendicular and parallel to the polarization plane, respectively.
The directivity originates from the combination of the time-periodic driving by light and the lattice symmetries of chiral semimetals.
These findings advance the understanding of nonlinear optical responses in chiral systems and open new avenues for next-generation optoelectronic applications.

The paper is organized as follows. In Sec.~\ref{sec:formalism}, we introduce the model Hamiltonian describing the electronic structure of CoSi under laser driving and the Floquet-Keldysh formalism for calculating the photocurrent. In Sec.~\ref{sec:results}, we present the numerical results and explain the detailed behavior of the dc photocurrent and HHG. In Sec.~\ref{sec:discussion}, we briefly discuss the selection rules for HHG, the experimental feasibility of our setup, and the effects of spin-orbit coupling. Our conclusions are given in Sec.~\ref{sec:conclusion}.

\section{Formalism}
\label{sec:formalism}

We start with a four-band tight-binding model to describe the electronic structure of CoSi \cite{Chang2017,Flicker2018}. The model qualitatively reproduces the band structure of CoSi near the Fermi level and provides a reasonable starting point for capturing the essential features of B20-type chiral semimetals. The model Hamiltonian is given by
\begin{align}
    H(t)=
    \sum_{j,X,k,Y}
    h_{jX,kY}e^{-i(e/\hbar)\vb*{A}(t)\cdot\vb*{\delta}_{jX,kY}}
    c_{j X}^\dagger c_{kY},
\end{align}
where $c_{j X}^{(\dagger)}$ denotes the annihilation (creation) operator of an electron at the $X$ sublattice of the $j$th site, $h_{jX,kY}$ represents a transfer integral, and $\vb*{\delta}_{jX,kY}$ represents the vector from the $Y$ sublattice of the $k$th site to the $X$ sublattice of the $j$th site.
The coordinates of each sublattice are $(0,0,0)$, $(a/2,a/2,0)$, $(a/2,0,a/2)$, and $(0,a/2,a/2)$, where the lattice constant is $a=0.44~\mathrm{nm}$.
The effect of the light irradiation is included through the Peierls phase.
The vector potential is given by $\vb*{A}(t)=(E/\Omega)(\sin\Omega t,\cos\Omega t,0)$, where $E$ and $\Omega$ are the amplitude and frequency of the light electric field, respectively.
In the above model, spin-orbit coupling is neglected, and the spin degree of freedom simply doubles the photocurrent.

We consider the nearest-neighbor and next-nearest-neighbor hopping processes, where their transfer integrals are schematically illustrated in Fig.~\ref{fig_system}(b).
The nearest-neighbor transfer integral is given by $h_{\pm} = h_1 \pm h_c$, where the sign depends on the bond.
The blue and orange bonds correspond to the positive and negative sign, respectively.
A finite $h_{\mathrm{c}}$ introduces chirality into the model. The next-nearest-neighbor transfer integral is given by $h_2$.
The transfer integrals are set as $h_1 = 0.34~\mathrm{eV}$, $h_{\mathrm{c}} = 0.14~\mathrm{eV}$, and $h_2 = 0.13~\mathrm{eV}$ to reproduce DFT bands of CoSi. Additionally, the on-site energy is set to $h_0 = 0.60~\mathrm{eV}$ to appropriately tune the Fermi level.
The energy band structure without light irradiation along the path in the Brillouin zone shown in Fig.~\ref{fig_system}(c) is presented in Fig.~\ref{fig_system}(d). The spin-1 chiral fermion is observed at the $\Gamma$ point, and the double Weyl fermion at the R point \cite{Tang2017,Chang2017}.
The band structure is consistent with the DFT calculations \cite{Tang2017,Xu2020}.

We use the Floquet-Keldysh formalism to describe the nonequilibrium states driven by light irradiation \cite{Tsuji2008,Tsuji2009,Aoki2014}. We assume that a time-periodic nonequilibrium state is realized through interactions with a reservoir.
The single particle Green's function is given by $G_{\vk,XY}(t,t^\prime)=-(i/\hbar)\langle T_C c_{\vk X,\mathrm{H}}(t)c_{\vk Y,\mathrm{H}}^\dagger(t^\prime)\rangle$.
Using the Floquet Green's function, the Dyson equation is written as
\begin{align}
    \begin{pmatrix}
        G^{\mathrm{R}}_{\vk\omega} & G^{\mathrm{K}}_{\vk\omega} \\
        0 & G^{\mathrm{A}}_{\vk\omega}
    \end{pmatrix}^{-1}
    =
    \begin{pmatrix}
        G^{0\mathrm{R}}_{\vk\omega} & 0 \\
        0 & G^{0\mathrm{A}}_{\vk\omega}
    \end{pmatrix}^{-1}
    +
    \begin{pmatrix}
        i\gamma & 2i\gamma F_\omega \\
        0 & -i\gamma
    \end{pmatrix},
\end{align}
where $\gamma$ characterizes the coupling strength between the system and the reservoir, $(F_\omega)_{XY,ml}=\tanh[\beta(\ho+m\hO)/2]\delta_{X,Y}\delta_{m,l}$ determines the form of the nonequilibrium distribution, and $\beta$ is the reservoir inverse temperature.
The above treatment is a minimal method that incorporates the effect of energy dissipation from the system to the reservoir by introducing two phenomenological parameters $\gamma$ and $\beta$.

The photocurrent is time periodic as a consequence of the above assumption and can be expressed as $\expval{J_\alpha(t)}=\sum_n e^{-in\Omega t}\expval{J_{\alpha,n}}$ ($\alpha=x,y,z$).
The Fourier coefficient of the $n$th harmonic is given by
\begin{align}
    \expval{J_{\alpha,n}}
    =2ie\hbar
    \int_{\mathrm{BZ}} \frac{d\vk}{(2\pi)^3}
    \int_{-\Omega/2}^{\Omega/2}\frac{d\omega}{2\pi}
    \mathrm{Tr}
    \qty[
        v_{\alpha\vk}
        S_n
        G^<_{\vk\omega}
    ], \label{eq_photocurrent}
\end{align}
where $\mathrm{Tr}[\cdot]$ denotes the trace over the photon-number indices, $v_{\alpha\vk} := \hbar^{-1} \partial_{k_\alpha} \mathcal{H}_{\vk}$ is the velocity operator defined as the derivative of the Floquet Hamiltonian $\mathcal{H}_{\vk}$ with respect to the wave vector $k_\alpha$, and $(S_n)_{XY,ml}=\delta_{X,Y}\delta_{m,l-n}$ originates from the phase factor $e^{-in\Omega t}$.
The prefactor 2 accounts for the spin degree of freedom.
It is worth noting the $E$ dependence of $\expval{J_{\alpha,n}}$.
The leading term of $(v_{\alpha\vk})_{ml}$ and $(G^<_{\vk\omega})_{ml}$ is of order $E^{|m-l|}$, so that one can expect $\expval{J_{\alpha,n}}\propto E^n$ when $eEa/(\hbar\Omega)\ll1$.
The harmonic components of the photocurrent give rise to HHG. The HHG intensity is given by
\begin{align}
    &I(n)=I_{\parallel}(n)+I_{\perp}(n), \\
    &I_{\parallel}(n)=\abs{n\Omega J_{\parallel,n}}^2,~
    I_{\perp}(n)=\abs{n\Omega J_{\perp,n}}^2,
\end{align}
where we define the in-plane and out-of-plane photocurrents as
$J_{\parallel}(t):=J_{x}(t)+iJ_{y}(t)$ and $J_{\perp}(t):=J_{z}(t)$, respectively.
The HHG intensity is decomposed into components parallel and perpendicular to the polarization plane. Note that $J_{\parallel,n}$ induces HHG perpendicular to the polarization plane, and $J_{\perp,n}$ induces HHG within the polarization plane.

\begin{figure}[t]
\begin{center}
\includegraphics[width=1\hsize]{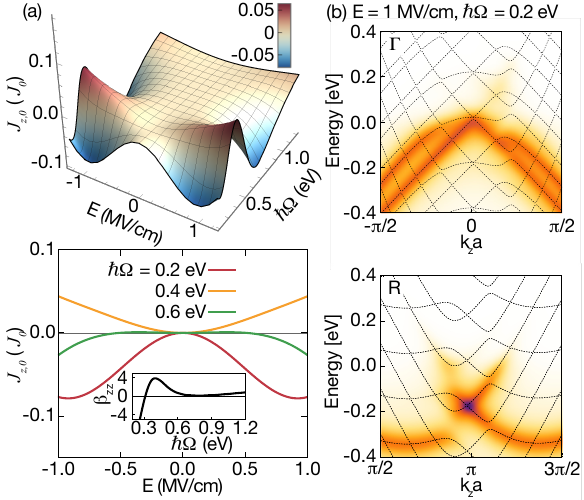}
\end{center}
\caption{(a) DC photocurrrent $J_{z,0}$ in plane of $E$ and $\Omega$ (upper panel) and those for various values of $\Omega$ as a function of $E$ (lower panel) for a narrower window of light intensity $\abs{E} \le 1~\mathrm{MV/cm}$. Here, the unit is $J_0=\frac{2}{(2\pi)^4}\frac{e^2}{\hbar a^2}\mathrm{V}\simeq 1.6\times10^{12}\mathrm{A/m^2}$. Inset shows the coefficient $\beta_{zz}$.
(b) Nonequilibrium electron distribution spectrum $N_{\vk}(\omega)$ across $\Gamma$ point (upper panel) and that across R point (lower panel) in the $k_z$ direction for a relatively weaker light intensity of $E=1~\mathrm{MV/cm}$. The dashed curves indicate the Floquet bands. We set $\gamma = 0.05~\mathrm{eV}$ and $\beta^{-1} = 0.03~\mathrm{eV}$.
}
\label{fig_pc1}
\end{figure}

\begin{figure}[t]
\begin{center}
\includegraphics[width=1\hsize]{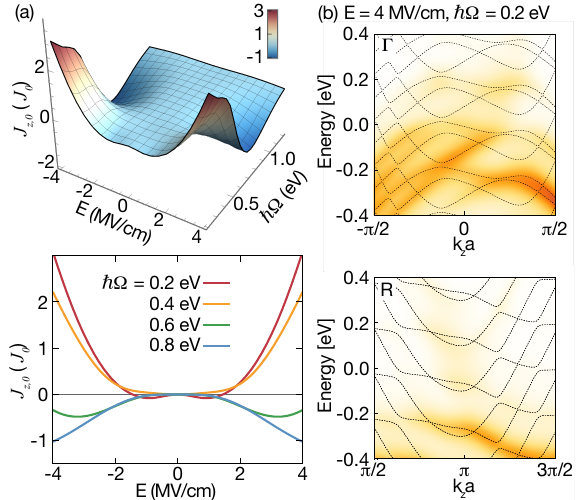}
\end{center}
\caption{(a) DC photocurrrent $J_{z,0}$ in plane of $E$ and $\Omega$ (upper panel) and those for various values of $\Omega$ as a function of $E$ (lower panel) for a larger window of light intensity $\abs{E} \le 4~\mathrm{MV/cm}$. The unit is $J_0=\frac{2}{(2\pi)^4}\frac{e^2}{\hbar a^2}\mathrm{V}\simeq 1.6\times10^{12}\mathrm{A/m^2}$.
(b) Nonequilibrium electron distribution spectrum $N_{\vk}(\omega)$ across $\Gamma$ point (upper panel) and that across R point (lower panel) in the $k_z$ direction with for a relatively stronger light intensity of $E=4~\mathrm{MV/cm}$. The dashed curves indicate the Floquet bands. We set $\gamma = 0.05~\mathrm{eV}$ and $\beta^{-1} = 0.03~\mathrm{eV}$.
}
\label{fig_pc2}
\end{figure}

\begin{figure}[t]
\begin{center}
\includegraphics[width=1\hsize]{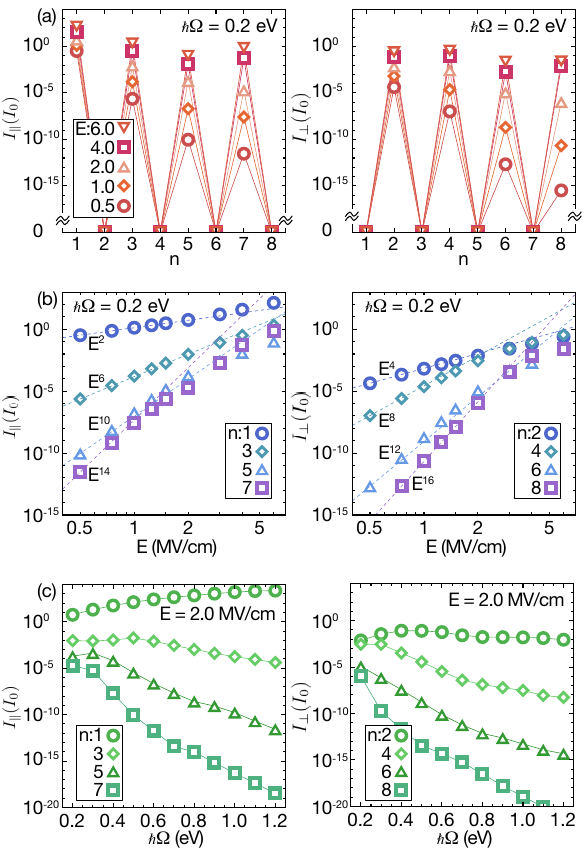}
\end{center}
\caption{
(a) HHG intensity $I_{\parallel}$ and $I_{\perp}$ as a function of $n$ for several values of $E$.
The unit is given by $I_0=(J_0\hbar^{-1}~\mathrm{eV})^2$.
(b) $I_{\parallel}$ and $I_{\perp}$ as functions of $E$ for odd and even $n$, respectively.
(c) $I_{\parallel}$ and $I_{\perp}$ as functions of $\hbar\Omega$ for odd and even $n$, respectively.
We set $\gamma = 0.05~\mathrm{eV}$ and $\beta^{-1} = 0.03~\mathrm{eV}$.}
\label{fig_hhg}
\end{figure}

\section{Results}
\label{sec:results}

In our setup, both the dc photocurrent and HHG arise. First, we show the results of the dc photocurrent in Figs.~\ref{fig_pc1} and \ref{fig_pc2}, and then show the results of HHG intensity in Fig.~\ref{fig_hhg}.
The photon number is restricted to $|m|\le10$ in all numerical calculations, and Eq.~(\ref{eq_photocurrent}) is discretized and numerically integrated over wavenumber $\vk$ and frequency $\omega$ using a $40^3 \times 60$ mesh. Under these conditions, the numerical results show sufficient convergence, ensuring that the qualitative features remain unchanged and further increasing the mesh resolution leads to only minor quantitative corrections.

Figure \ref{fig_pc1}(a) shows the dc photocurrent $J_{z,0}$ as a function of the electric field $E$ and the frequency $\Omega$. The bottom panel shows the corresponding $J_{z,0}$ curves for specific values of $\Omega$.
For a weak $E$-field ($|E|\ll1~\mathrm{MV/cm}$), the photocurrent exhibits a quadratic dependence on $E$, described as $J_{z,0} = \beta_{zz} \tau E^2$, where $\tau = \hbar / 2\gamma$ is the relaxation time. The coefficient $\beta_{zz}$ is expected to be quantized under appropriate conditions \cite{De_Juan2017-yl,Flicker2018,Orenstein2021}.
Here, we evaluate the coefficient $\beta_{zz}$ by fitting the calculated photocurrent $J_{z,0}$ in the weak $E$-field regime to a quadratic curve. The coefficient $\beta_{zz}$ is shown in the inset of the bottom panel. $\beta_{zz}$ changes its sign from negative to positive around $\hO=0.3~\mathrm{eV}$, and reaches a maximum near $\hO=0.4~\mathrm{eV}$.
Under the current conditions, the quantization of $\beta_{zz}$ is not observed, consistent with previous numerical simulations \cite{Ni2020,Ni2021}.
As the $E$-field increases, higher-order corrections become significant, leading to deviations from the quadratic approximation.
Above $E \simeq 0.5~\mathrm{MV/cm}$, $J_{z,0}$ as a function of $\Omega$ exhibits multiple sign changes, where higher-order terms in powers of $E$ play a significant role in determining the sign of $J_{z,0}$.
Note that the sign of $J_{z,0}$ can be reversed by changing the handedness of circularly polarized light, reflecting the chirality of CoSi.

In our formulation, the dc photocurrent generation is attributed to asymmetric Floquet bands along the $k_z$ direction and an asymmetric nonequilibrium distribution.
Figure \ref{fig_pc1}(b) shows the nonequilibrium electron distribution spectrum
\begin{align}
    N_{\vk\omega}
        =&\int_{0}^{T}\frac{d\ta}{T}
        \int_{-\infty}^{\infty}d\tr e^{i\omega\tr} \notag \\
        &\times\qty(-\frac{i\hbar}{2\pi})\sum_XG^<_{\vk,XX}(\ta+\tr/2,\ta-\tr/2),
\end{align}
where the averaged time $\ta=(t+t^\prime)/2$ and the relative time $\tr=t-t^\prime$ are introduced.
The top and bottom panels show the spectrum $N_{\vk\omega}$ across the $\Gamma$ and R points, respectively, along the $k_z$ direction.
Because of the chirality of the system, the spectrum shows asymmetric distribution, reflecting the asymmetric Floquet bands and nonequilibrium distribution function. The asymmetric spectrum leads to the dc photocurrent generation.

Figure \ref{fig_pc2}(a) shows the photocurrent $J_{z,0}$, including the strong $E$-field regime $(|E|\gtrsim1~\mathrm{MV/cm})$. At relatively low frequencies $\Omega\lesssim0.5~\mathrm{eV}$, $J_{z,0}$ is positive and increases with the $E$-field, whereas at higher frequencies $\Omega\gtrsim0.5~\mathrm{eV}$, $J_{z,0}$ becomes negative. In this regime, the nonlinearity becomes more pronounced, and the electronic structure deviates significantly from equilibrium.
Figure \ref{fig_pc2}(b) shows the spectrum $N_{\vk\omega}$ under the intense $E$-field condition. The spectrum reveal highly asymmetric features, reflecting the hybridization of photon-dressed bands. The spectrum is broadened and widely distributed, further reflecting the strongly driven nonequilibrium nature of the chiral system.

Figure \ref{fig_hhg}(a) shows semilog plots of the HHG intensity $I_\parallel$ and $I_\perp$, respectively, as a function of the harmonic order $n$ for several values of $E$. The HHG intensity increases with increasing $E$.
One can see that $I_\parallel$ is finite only for odd harmonic orders, while $I_\perp$ is finite only for even harmonic orders.
Consequently, the HHG intensity of odd-harmonic orders exhibits perpendicular directivity relative to the polarization plane, whereas the HHG intensity of even-harmonic orders exhibits directivity within the polarization plane. The directivity of the HHG intensity, which depends on the parity of the harmonic order, is a characteristic feature of chiral semimetals.

The parity-dependent directivity of HHG can be understood based on dynamical symmetry \cite{Simon1968,Tang1971,alon1998,Saito2017,Neufeld2019,ikeda2019,ikeda2020,kanega2021,kanega2024}.
From the combination of the screw rotation symmetry of B20-type materials, $s_{2z}=\{C_{2z}|\frac{1}{2} 0 \frac{1}{2}\}$, and the time periodicity of the photocurrent $J_{\alpha}(t+T)=J_{\alpha}(t)$, we obtain $J_{\parallel}(t+T/2)=e^{-i\pi}J_{\parallel}(t)$ and $J_{\perp}(t+T/2)=J_{\perp}(t)$. These relations result in the following selection rules for HHG: $I_{\parallel}(n) = 0$ except for $n \equiv 1 \pmod{2}$ and $I_{\perp}(n) = 0$ except for $n \equiv 0 \pmod{2}$.

To clarify the $E$-field dependence of HHG in more detail, Figure~\ref{fig_hhg}(b) shows log-log plots of the HHG intensity as a function of $E$.
In the region where the $E$-field is relatively weak ($E\lesssim1~\mathrm{MV/cm}$), the $n$th order HHG intensity is proportional to $E^{2n}$. This behavior is consistent with the $n$th order response of $J_{\alpha,n}(\propto E^n)$ in the perturbative regime.
As the $E$-field increases, the $E^{2n}$ lines intersect in $E\gtrsim1~\mathrm{MV/cm}$, indicating the breakdown of perturbative picture. In the strong $E$-field regime, non-perturbative effects become significant. These effects include highly hybridized photon-dressed bands and contributions from high-energy photon-dressed states. As a result, deviations from the $E^{2n}$ lines are observed.

Finally, the frequency $\Omega$ dependence of the HHG intensity is explained. Figure \ref{fig_hhg}(c) shows semilog plots of the HHG intensity as a function of $\Omega$ for the fixed value of $E=2.0~\mathrm{MV/cm}$.
For $n=1$, $I_\parallel$ increases monotonically with $\Omega$. For $n>2$, the HHG intensity tends to decrease as $\Omega$ increases. Furthermore, as $n$ increases, the reduction in HHG intensity with increasing $\Omega$ becomes more pronounced. This suggests that HHG detection is more feasible at lower frequencies.

\section{Discussion}
\label{sec:discussion}

The selection rules discussed above can be extended to systems with an $N$-fold rotational or screw symmetry axis. When light propagates along either of these axes, the in-plane and out-of-plane photocurrents satisfy the following conditions:
\begin{align}
    &J_{\parallel}(t+T/N)=e^{-i2\pi/N}J_{\parallel}(t), \\
    &J_{\perp}(t+T/N)=J_{\perp}(t).
\end{align}
From these conditions, the following selection rules for HHG are obtained
\begin{align}
    &I_{\parallel}(n)=0\quad\mathrm{except~for}\quad n\equiv\pm1\pmod{N}, \\
    &I_{\perp}(n)=0\quad\mathrm{except~for}\quad n\equiv0\pmod{N}.
\end{align}
Thus, our results indicate that the HHG spectrum can serve as a probe for detecting lattice structures that break spatial inversion symmetry and preserving an $N$-fold rotational or screw symmetry axis.

The nonlinear optical responses of B20-type chiral semimetals have attracted significant interest, with several experimental groups investigating the detection of the circular photogalvanic effect \cite{Rees2020,Ni2020,Ni2021}.
Recent advances in high-intensity laser technology have facilitated experimental studies on the nonlinear optical responses of topological materials \cite{Wu2017,Patankar2018,Osterhoudt2019,Sirica2019}.
In particular, HHG in topological insulators and Weyl semimetals has been experimentally observed by several groups at electric field strengths comparable to or exceeding those considered in this work \cite{bai2021high,Schmid2021,Lv2021,Baykusheva2021,Heide2022}.
As proposed in this study, measuring the dc photocurrent beyond the second-order regime and observing parity-dependent directivity of HHG provide valuable opportunities to explore the nonequilibrium properties of chiral semimetals.

It is known that incorporating spin-orbit coupling modifies the energy bands and alters the topological properties of chiral fermions \cite{Tang2017,Chang2017,Pshenay-Severin2018}. Such band modifications are not expected to affect the qualitative features of the main results of this work, whereas the inclusion of spin-orbit coupling is expected to improve the accuracy of the theoretical estimation of the photocurrent, as confirmed in previous numerical simulations on optical conductivities \cite{Xu2020,Ni2020,Ni2021}.

\section{Conclusion}
\label{sec:conclusion}

We demonstrate the bulk dc photocurrent and high-harmonic generation (HHG) induced by circularly polarized light in B20-type chiral semimetals.
The dc photocurrent exhibits multiple sign changes depending on the light electric field $E$ and the frequency $\Omega$, reflecting the nonreciprocal excitation of chiral fermions and the underlying chirality of the lattice structure.
For HHG, the directivity of the HHG intensity depends on the parity of the harmonic orders. Namely, the HHG intensity of odd-order harmonics exhibits directivity perpendicular relative to the polarization plane, while that of even-order harmonics exhibits directivity within the polarization plane.
The parity-dependent directivity of HHG, along with the appearance of even-harmonic generation, is the distinctive feature of chiral semimetals.
This study provides a foundation for exploring nonlinear optical responses induced by circularly polarized light and offers insights into the nonequilibrium properties of chiral semimetals, paving the way for future experimental and theoretical advancements in the strongly driven chiral systems.

\section*{Acknowledgments}

Y.O. is grateful to A. Yamakage for valuable discussions. This work was supported by JSPS KAKENHI (Grants No. JP20H00337, No. JP24H02231, and No. JP25H00611), JST CREST (Grant No. JPMJCR20T1), and Waseda University Grant for Special Research Projects (Grants No. 2024C-153, No. 2024C-290, and No. 2025C-133).

\end{document}